\documentstyle[twocolumn,aps,prl,floats,graphicx]{revtex}

\begin{document}

\wideabs{						

\title{
 Dynamic Response of Adhesion Complexes:
 Beyond the Single-Path Picture
}
\author{
 Denis Bartolo$^{1,}$\cite{denis},
 Imre Der\'enyi$^{2,}$\cite{imre}, and
 Armand Ajdari$^{1,}$\cite{armand}
}
\address{
 $^{1}$Laboratoire de Physico-Chimie Th\'eorique, esa 7083 CNRS, ESPCI,
  10 rue Vauquelin, F-75231 Paris C\'edex 05, France \\
 $^{2}$Institut Curie, UMR 168,
  26 rue d'Ulm, F-75248 Paris C\'edex 05, France
}
\date{\today}

\maketitle

\begin{abstract}
~							
We analyze the response of molecular adhesion complexes to increasing
pulling forces (dynamic force spectroscopy) when dissociation can
occur along either one of two alternative trajectories in the underlying
multidimensional energy landscape. A great diversity of behaviors
(e.g. non-monotonicity) is found for the unbinding force and time as a
function of the rate at which the pulling force is increased. We
highlight an intrinsic difficulty in unambiguously determining the
features of the energy landscape from single-molecule pulling
experiments. We also suggest a class of ``harpoon'' stickers that bind
easily but resist strong pulling efficiently.
\end{abstract}

\draft


}							

\begin{figure}[!b]
\centerline{\includegraphics[scale=1]{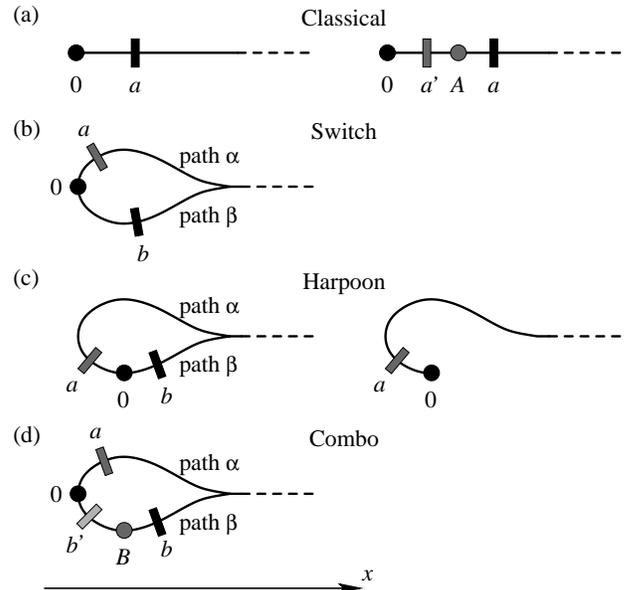}}
\caption{
Sketch of the topology of the main valley of the energy landscape
for a few examples.
$0$ denotes the fundamental bound state, $A$ and $B$ are local minima,
and $a$, $a'$, $b$ and $b'$ are passes to overcome. To the right (increasing values
of $x$) of the last passes is the continuum that describes unbound states.
(a) classical single-path scheme.
(b,c,d) unbinding can occur through two alternative routes
$\alpha$ and $\beta$.
}\label{f_1}\end{figure}

\begin{figure}[!t]
\centerline{\includegraphics[scale=1]{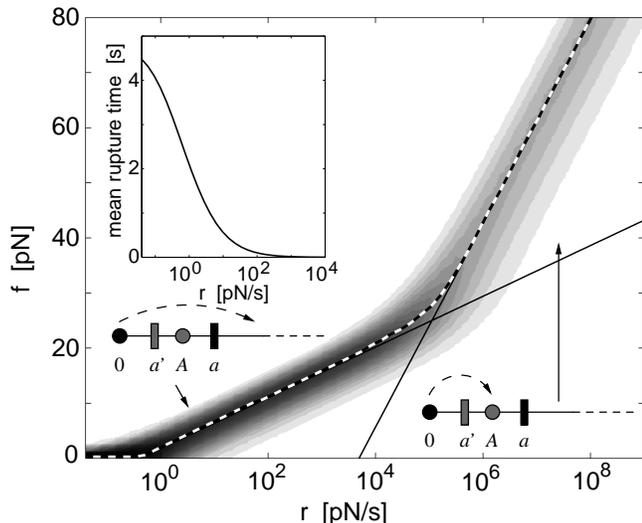}}
\caption{
Classical picture for a single-path energy
landscape~\protect\cite{Eva} (Fig.~\ref{f_1}a):
the probability density $P(f)$ for unbinding  at force $f$
is plotted in grey-scale as a function of the pulling rate $r$.
The typical force $f_{\rm typ}$ (locus of the maximum of $P$) is
highlighted with a dashed-line.
Plotted curves correspond to
$E_{a'}=12$, $x_{a'}=0.5$,
$E_A=9$, $x_A=1$, and
$E_a=20$, $x_a=2$.
At very low pulling rates unbinding is not affected by the pulling and proceeds
over barrier $a$ with a ``spontaneous'' rate $\omega_0 \exp(-E_a)$. For larger pulling
rates the typical unbinding force $f_{\rm typ}$ increases linearly with
$\log(r)$, with
a slope proportional to $1/x_a$. Increasing
further the pulling rate can lead to a steeper slope $\propto 1/x_{a'}$ corresponding to escape
over the inner barrier $a'$.
These asymptotes are depicted with solid lines. The dashed arrows along
the drawings indicate which pairs of energy well and barrier
are probed in these asymptotic limits.
Inset: mean rupture time against pulling rate.
}\label{f_2}\end{figure}

The last decades have witnessed a remarkable development
of physical investigation methods to probe single molecules or
complexes by various micromanipulation means.
New techniques have been put forward to probe
the unfolding of proteins and to quantify the strength
of adhesion structures~\cite{titin,Poi01,Sim99,Nis00,Pie96}.
An important step in this direction
is the proposal of the group of Evans
to use soft structures to pull on adhesion complexes
or molecules at various loading rates (dynamic force spectroscopy)~\cite{Eva}.
Moving the other end of the soft structure at constant velocity
induces on the complex a pulling force that
increases linearly in time $f=rt$.
Measuring the typical rupture time $t_{\rm typ}$
yields a typical rupture force $f_{\rm typ}=rt_{\rm typ}$ that depends
on the pulling rate $r$.
This provides information as to the energy landscape
of the bound complex.
Indeed, in many situations one observes a linear increase of $f_{\rm typ}$
with $\log(r)$, which can be understood within  a simple adiabatic Kramers picture
for the escape from a well (bound/attached state)
over a barrier of height $E$ located at a projected distance $x$
from the well along the pulling direction. The progressive increase
of the force
results in a corresponding increase of the escape rate,
so that, in agreement with some experiments~\cite{Eva},
the typical rupture force {\em increases logarithmically with $r$}:
$f_{\rm typ} \simeq k_{\rm B}T/x \ln [rx/(k_{\rm B}T\omega)]$,
where $\omega$ is the escape rate in the absence of force.
The rupture time on the other hand {\em decreases with $r$}.
The occurrence in some cases of two successive straight lines in a
$[f_{\rm typ},\log(r)]$ plot
has been argued to be the consequence of having two
successive barriers along the 1D escape path,
the intermediate one showing up in the response at fast pulling rates~\cite{Eva}
(Figs.~\ref{f_1}a and \ref{f_2}).
Other theories have tried to back up more complete
information as to the overall effective 1D potential landscape by an analysis
of the probability distribution for rupture time and of the statistics
of trajectories before rupture~\cite{Hum01,Hey00}. Assemblies in series and in parallel
of such 1D bonds have also been considered~\cite{Eva01,Sei00,Gau98a}.

\begin{figure}[!t]
\centerline{\includegraphics[scale=1]{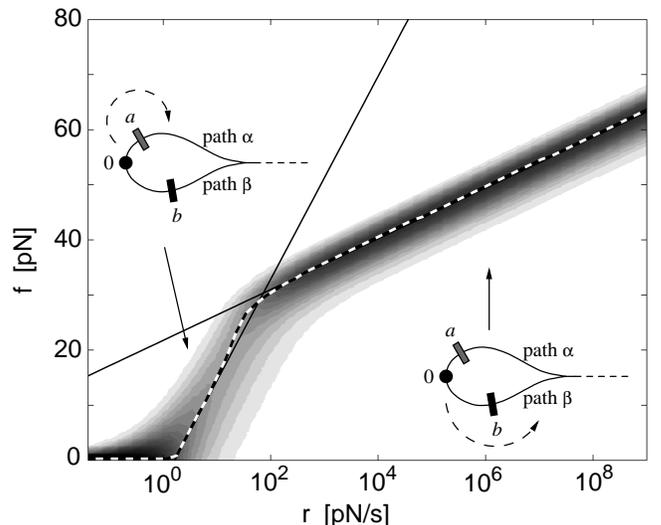}}
\caption{
Switch geometry (Fig.~\ref{f_1}b): plot of the same quantities as in
Figure \ref{f_2}, for
$E_a=20$, $x_a=0.5$, and
$E_b=30$, $x_b=2$.
At low pulling rates unbinding is controlled by escapes over $a$
whereas for large values of $r$ it occurs mostly over $b$:
the slope of the unbinding force (average or typical)
decreases from $1/x_a$ to $1/x_b$.
}\label{f_3}\end{figure}

\begin{figure}[!t]
\centerline{\includegraphics[scale=1]{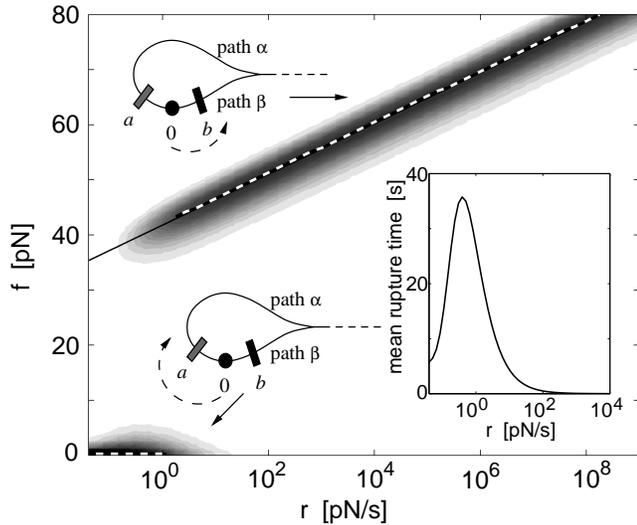}}
\caption{
``Harpoon'' geometry (Fig.~\ref{f_1}c): plot of the same quantities as in
Figure \ref{f_2}, for
$E_a=20$, $x_a=-2$, and
$E_b=40$, $x_b=2$.
Pulling here impedes unbinding through the ``spontaneous'' route $\alpha$,
so that as soon as the rate is strong enough for pulling to affect
unbinding, the escape is controlled by the
larger barrier $b$, resulting in an upward jump of the typical
unbinding force and time. Inset: the average unbinding time is here
non-monotonic.
}\label{f_4}\end{figure}

\begin{figure}[!t]
\centerline{\includegraphics[scale=1]{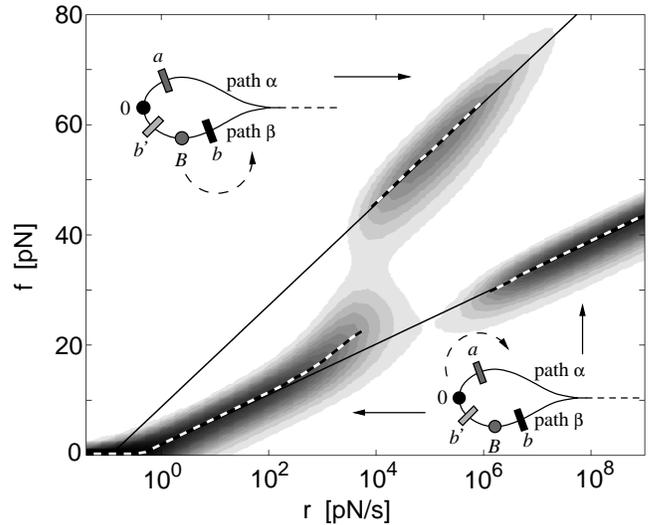}}
\caption{
``Selective harpoon'' from the combo topology (Fig.~\ref{f_1}d):
same quantities as in Figure \ref{f_2}, for
$E_a=20$, $x_a=2$,
$E_{b'}=10$, $x_{b'}=0.5$,
$E_B=5$, $x_B=1.5$, and
$E_{b}=27$, $x_{b}=2.5$.
At low pulling rates the spontaneous path $\alpha$ is used. Upon increase
of $r$, larger forces are employed and the minimum $B$ becomes favorable
as compared to $0$.
As $E_{b'}$ is not too large,
equilibration of population then empties $0$ in $B$,
so that escape eventually occurs from $B$ over $b$,  resulting in a higher straight line of slope
$\propto 1/(x_{b}-x_B)$. At even higher pulling rates,
because $x_a > x_{b'}$, the escape over $a$ becomes faster than
this equilibration, and therefore, path $\alpha$ is used again.
Barrier $a$ controls
the behavior at low and high rates, but in an intermediate window,
a stronger bonding is provided by barrier $b$.
The typical (dashed line) or average unbinding force is non-monotonic.
}\label{f_5}\end{figure}

\begin{figure}[!t]
\centerline{\includegraphics[scale=1]{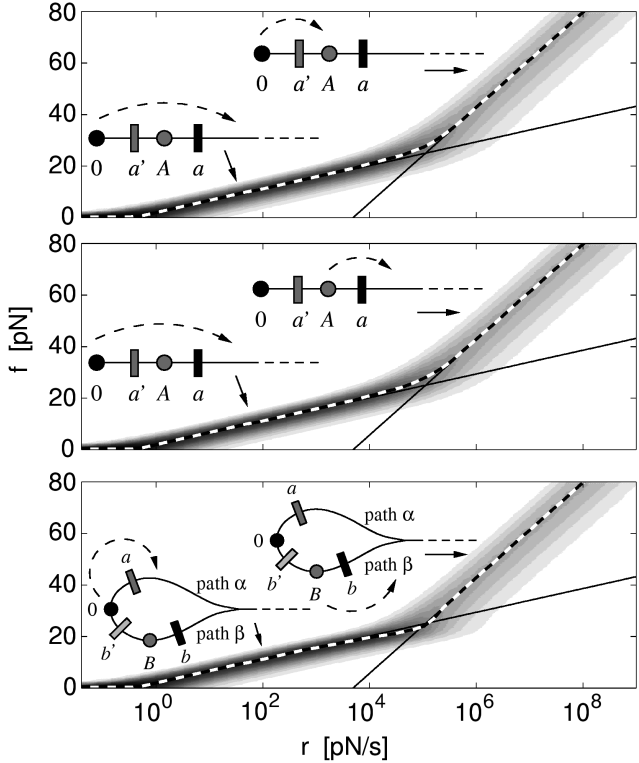}}
\caption{
Similar curves
obtained from significantly different energy landscapes.
(a) classical single-path of Figure \ref{f_1}a, data of Figure \ref{f_2}.
(b) classical single-path of Figure \ref{f_1}a, with
$E_{a'}=11$, $x_{a'}=1$,
$E_A=8$, $x_A=1.5$, and
$E_a=20$, $x_a=2$.
(c) combo two-path geometry (Fig. \ref{f_1}d), with
$E_a=20$, $x_a=2$,
$E_{b'}=18$, $x_{b'}=2.5$,
$E_B=15$, $x_B=3$, and
$E_{b}=27$, $x_{b}=3.5$.
In all cases the straight part for weak $r$ corresponds to
escape over $a$ from $0$. The second steeper slope corresponds to
escape over $a'$ from $0$ in case (a), over $a$ from $A$ in case (b),
over $b$ from $B$ in case (c).
Neither the topology, nor the location of the probed segment of the
energy landscape can be asserted from such data sets.
}\label{f_6}\end{figure}

In this Letter we point out limitations
arising from the a priori assumption
of a single-path topology of the energy landscape  for the interpretation of such
experiments. From the analysis of simple examples with a two-path topology,
we draw three conclusions:
(i) first, the dependence of the rupture force and rupture time  on the pulling rate can take various forms,
including {\em non-monotonic} behavior
(see e.g. Figs.~\ref{f_3} to \ref{f_5}).
(ii) Second, the main features of the energy landscape
can not be unambiguously deduced from a [$f_{\rm typ},\log(r)$] plot,
as very different landscapes can yield similar curves
(Fig.~\ref{f_6}).
(iii) Third, we propose simple ``harpoon'' designs (Fig.~\ref{f_1} c and d)
for functionally efficient stickers
that can bind easily but resist strongly
in a range of pulling forces (Fig.~\ref{f_4}).

Obviously for real binding/adhesion complexes, there are numerous (conformational)
degrees of freedom, and the configurational space is clearly multidimensional.
This allows for complex energy landscapes
and various topologies for the structure of their valleys and passes~\cite{Pac00}.
Only the probing (pulling) is unidirectional.
We note in passing that even for more macroscopic sticky systems,
usual adhesion tests for soft adhesives often show up hysteresis loops associated
with the existence of more than one degree of freedom \cite{JKR71}.
We do not attempt here an exhaustive
exploration of effects allowed by the multidimensionality of the
phase space,
but rather focus on a few simple two-path topologies (Fig.~\ref{f_1}),
to argue for the three points mentioned above.

The three examples we consider, sketched in Figure \ref{f_1} b, c, and d, 
correspond to simple hairpin schemes
whereby detachment can proceed through two alternative
routes $\alpha$ and $\beta$.
These simple quasi 1D schematic situations can be conveniently
dealt with using an adiabatic Kramers theory, which has 
been shown to be an efficient way of obtaining semi-quantitatively 
correct answers \cite{Sei98}.

A common set of notations can be ascribed for all cases
(Fig.~\ref{f_1}). From the fundamental bound state ``0'', the route $\alpha$ for
escape (detachment) is over barriers $a$, of height $E_{a}$
located at a projected distance $x_{a}$ from ``0''. Alternatively, 
escape can occur through branch $\beta$, over barrier $b$,
of height $E_{b}$ and projected distance $x_{b}$.
All energies and projected distances are measured relative to the state ``0''
(i.e. $E_0=0$ and $x_0=0$). Intermediate barriers $a'$, $b'$ and local minima A and B
may exist, with energies
$E_{a'}, E_{b'}, E_A, E_B$ (all positive), and projected distances
$x_{a'}, x_{b'}, x_A, x_B$. In line with typical values from
experiments, we choose to write energies in units of
$k_{\rm B}T\simeq 4$~pN$\,$nm and
distances in nm.

Practically, we describe the time evolution of the probabilities
of being in the potential minima (bound states) using ``chemical''
transition rates over the barriers as given by the Kramers formula.
We furthermore assume the attempt frequencies to be constant and all equal to $\omega_0$
which provides the only intrinsic time-scale in the problem,
so that the transition rate from minimum $I$ over the neighboring barrier $i$
is $\omega_0 \exp[-(E_i-E_I) + f(t)(x_i-x_I)]$.
For the plots of Figures \ref{f_2} to \ref{f_6} we take arbitrarily
$\omega_0=10^8$~s$^{-1}$.
Jump over the rightmost barrier ($a$ or $b$) of either path
corresponds to rupture leading to escape to $x \rightarrow \infty$.

We focus on the case where either $E_{b'}$ or $E_{b}$ is larger
than $E_{a}$ so that $\alpha$ is the ``natural'' route by which attachment
and detachment proceeds {\em in the absence of pulling}.
We also limit ourselves to simple scenarios in which the force is linearly increased in time
$f=rt$.

For further reference we recall the classical single-path scenario
(Fig.~\ref{f_1}a) in Figure \ref{f_2} for a typical set of parameters,
and then we turn to a brief analysis of the three geometries we have
introduced (Fig.~\ref{f_1} b, c, and d).

{\em First case: switch} -- Topology as in Figure \ref{f_1}b.
Escape occurs through either barrier $a$ or barrier $b$ both located
downwards in the pulling direction ($x_a, x_b>0$).
The escape proceeds through path $\alpha$ at weak pulling rates as $E_a<E_b$,
but if $x_b>x_a$ it can switch to path $\beta$ for pulling forces $f$
large enough such that $E_a-f x_a >E_b- fx_b$.
The result (see Fig.~\ref{f_3}) is then a succession of two straight lines of decreasing slopes
in the [$f_{\rm typ}, \log(r)$] plot, the first one  (slope $\propto 1/x_a$)
characteristic of the spontaneous route $\alpha$
while the second (slope $\propto 1/x_b$) provides information on the
alternative route $\beta$.
In the trivial case $x_a>x_b$ route $\beta$ is never explored
so that the classical single-path picture applies.

To clarify the calculation leading to the plot in Figure \ref{f_3},
we describe the evolution of the probability of attachment $p(t)$ at time $t$
of the system initially attached at time $t=0$ [$p(0)=1$] by
\begin{equation}
\partial_t p(t) = - \omega_0 ( e^{-E_a+f(t)x_{a}} +
e^{-E_{b}+f(t)x_{b}} ) p(t)
\end{equation}
Solving (1) numerically with $f=rt$ yields $p(t)$ and therefore the
probability density
$P(f)=- \frac{1}{r}\, \partial_t p(\frac{f}{r})$ for the unbinding force.
The typical values of $f$ are highlighted in the plots,
with the whole distribution $P(f)$ suggested through a grey-scale.
Similar procedures will be used in the following examples, with thermal
equilibrium between the bound states assumed as initial conditions.

{\em Second case: harpoon} -- Topology similar to the previous one
but with $x_a<0$ (Fig.~\ref{f_1}c).
The main feature here is that
as the pulling force increases, the probability to escape over $a$ decreases.
Therefore the system
gets ``stuck'' in route $\beta$. If the barrier $E_b$ is infinite
(left side of Fig.~\ref{f_1}c),
there is a finite probability $p_{\infty}=\exp(-\frac{\omega_0 e^{-E_a}}{r|x_a|})$
that unbinding never occurs.  For a finite but high barrier $E_b$,
pulling eventually results in unbinding but at high rupture forces
(see Fig.~\ref{f_4}).
The topology thus allows here to form ``easily'' (i.e. over barrier $a$)
a ``harpoon'' sticker that can resist strong pulling.
Correspondingly the mean unbinding time increases first with pulling rate
(a phenomenology connected to the negative resistance analyzed in Ref.~\cite{Mag96}),
before decreasing for larger values when activated escape over $b$ dominates.
Note that the probability distribution $P(f)$, now consists of two separate ensembles,
which coexist over a narrow region of pulling rates.
This is in contrast with Figure \ref{f_3}
where there is a continuous evolution of a single cloud.

{\em Third case: combo} --
The alternative route consists of two barriers and a local minimum B
(Fig.~\ref{f_1}d),
and we focus on the case where $E_{b'}$ is smaller than the two others.

Thanks to the increased complexity and number of parameters in this case
many scenarios can occur, covering features already unveiled
in Figures \ref{f_2} to \ref{f_4} (e.g. switch and harpoon). More intricate pictures can also show up,
as depicted in Figure \ref{f_5}. An explanation of
this example is given in the caption, illuminating
how for low or high pulling rates barrier $a$ controls the behavior,
whereas for intermediate values, the secondary and stronger barrier $b$
limits unbinding.
Two features are striking. First, the unbinding force (typical or average) is no more
monotonic. Second, branch $\beta$ results in a strengthening of
the binding complex for a given window of pulling rates $r$ (selective harpoon).

{\em Discussion} --
With the three simple examples above, we have clearly enlarged the numbers of behaviors that
one may obtain from a classical dynamic force spectroscopy method
(see Figs. \ref{f_3} to \ref{f_5}).
Conversely, we also want to stress the second point (ii) mentioned in the introduction:
simple patterns (e.g. the succession of two lines of increasing slopes) can be the
outcome of many diverse landscapes.
For example, Figure \ref{f_6} displays force-rate curves similar to that of Figure \ref{f_2}, but that correspond to
sensibly different landscapes.
Not only are the typical and average unbinding forces very similar
in the three cases, but so are the probability distributions for most values of $r$.
Only close to the cross-over between the two straight lines can slight differences
be detected.
To distinguish
more selectively possible landscapes, it may be necessary to
use other temporal sequences than the simple $f=rt$, e.g. to reveal
equilibration processes between local minima.

Eventually we would like to emphasize
that the harpoon geometries proposed here constitute a very obvious
paradigm for efficient stickers.
Attachment of the sticker can proceed through route $\alpha$ with a possibly not too
high barrier $E_a$. The harpoon configuration then allows
to benefit from the much stronger $b$ barrier for a given window of
pulling forces, making the sticker more efficient in these conditions.
This ``hook'' design is obviously a favorable strategy for adhesion complexes,
the function of which is to maintain adhesion under the action of well-defined
tearing stresses. It would be surprising if advantage was not taken of this by
some biological systems.

\vspace{-7pt}						

\end{document}